\documentclass[12p]{article}
 \usepackage{dcolumn}

\usepackage{latexsym}
\usepackage{cite}
\usepackage{color}
\usepackage{amsmath}
\usepackage{epsfig}
\usepackage{hyperref}
\newcommand{\ortala}[1]{\begin{center}#1\end{center}}

\newcommand{\sandd}[1]{\left\langle #1\right\rangle}
\newcommand{\sanddr}[1]{\left\langle\left\langle #1\right\rangle\right\rangle_r}

\newcommand{\summ}[3]{{{\underset{#1 }{\overset{#2}{\displaystyle\sum}}}#3}}

\newcommand{\re}[1]{(\ref{#1})}

\newcommand{\eq}[2]{\begin{equation}\label{#1}  #2\end{equation}}

\newcommand{\paran}[1]{\left(#1\right)}

\newcommand{\sch}[1]{Schrodinger}

\newcommand{\komb}[2]{\paran{\begin{array}{c} #1 \\ #2 \end{array}}}

\linethickness{0.6pt}

\begin{document}

\ortala{\textbf{Hysteresis Behaviors of the Binary Ising Model}}

\ortala{\"Umit Ak\i nc\i\footnote{umit.akinci@deu.edu.tr}, G\"ul\c{s}en  Karakoyun }
\ortala{\textit{Department of Physics, Dokuz Eyl\"ul University,
TR-35160 Izmir, Turkey}}

\section{Abstract}\label{abstract}

Hysteresis behaviors of the binary alloy system represented by the formula $A_c B_{1-c}$ have been investigated within the framework of EFT. The system consists of type A atoms (spin-$1$) with concentration $c$ and type B atoms (spin-$1/2$) with concentration $1-c$. After giving the phase diagrams, we focused on the different type of hysteresis behaviors in the system. Especially, the mechanisms giving rise to double hysteresis behavior have been explained, which appear at large negative values of the crystal field and low temperatures. It has been observed that binary alloy system could exhibit DH behavior in region $0<c<0.557$. Besides, dependence of hysteresis loop area, coercive field and remanent magnetization on the concentration has been investigated.

\section{Introduction}\label{introduction}

Disordered binary alloys received attention due to their many physical realizations and rich critical behaviors.  
A typical binary alloy can be represented by $A_c B_{1-c}$, where $A$ and $B$ denote different type atoms which have concentrations $c$ and $1-c$, respectively. In  magnetic language, this means that atoms have different spins distributed on a lattice such that $c$ percentage of the lattice sites occupied by spin-$S_A$ and remaining sites have spin-$S_B$ atoms. These type of models can explain the magnetic systems such as 
$Rb_2 Mn_c Mg_{1 - c} F_4$ \cite{ref1},  $Mn _c Zn_{1 - c} F_2$ \cite{ref2} and  $Co(S_p Se_{1 - p})_2$ \cite{ref3}. Indeed binary alloys keep more space than these examples in the literature. Detailed investigation on the characteristics and importance of the transition metal alloys and rare earth alloys can be found in Ref.  \cite{ref4}.

On the theoretical side, literature contains works devoted to the critical and thermodynamical properties of the binary alloys such as Perturbation Theory \cite{ref5}, 
mean field approximation (MFA) \cite{ref6,ref7,ref8,ref9,ref10,ref11,ref12,ref13,ref14,ref15}, Bethe-Peirls Approximation \cite{ref16}, effective field theory (EFT) 
\cite{ref17,ref18,ref19,ref20,ref21,ref22,ref23,ref24,ref25,ref26,ref27}
and  Monte Carlo (MC) simulation \cite{ref28,ref29,ref30,ref31,ref32,ref33,ref34,ref35,ref36}. In theoretical literature, some other models exist such as Heisenberg model \cite{ref37}. Besides, binary alloys on some other geometries have been solved for superlattice \cite{ref38} and thin film \cite{ref39,ref40,ref41} structures.  Nowadays, hysteresis of these  type of geometries have attention also in an experimental point of view \cite{ref41_1,ref41_2}.

Recently it was demonstrated that, spin-1 Ising model could exhibit double hysteresis behavior according to the value of the crystal field \cite{ref42}. Later on, it was generalized to the spin-S model: spin -S Ising model displays $2S$ windowed hysteresis structure \cite{ref43}. Double  hysteresis behaviors are also observed experimentally in doped ceramics \cite{ref43_1} and superlattices \cite{ref43_2}.

This work can be considered as generalization of the investigation of the hysteresis behavior to the binary alloys. For this aim, the
paper is organized as follows: In Sec. \ref{model} we briefly
present the model and  formulation. The results and discussions are
presented in Sec. \ref{results}, and finally Sec. \ref{conclusion}
contains our conclusions.





\section{Model and Formulation}\label{model}

Binary system $A_cB_{1-c}$ consists of  type A (spin-1/2) and type B (spin-1) atoms, which are distributed on a regular lattice with coordination number $z$ such that c percentage of the atoms are the type of A and remaining are type of B.    
The Hamiltonian of this binary Ising model with uniform longitudinal
magnetic field is given by
\eq{denk1}{\mathcal{H}=-J\summ{<i,j>}{}{\paran{\xi_i \xi_j J_{AA} \sigma_i \sigma_j+
\xi_i \delta_j J_{AB} \sigma_i s_j+
\delta_i \xi_j J_{BA} s_i \sigma_j+
 \delta_i \delta_j J_{BB} s_i s_j}}-D\summ{i}{}{\delta_i s_i^2}-H\summ{i}{}{\paran{\xi_i \sigma_i + \delta_i s_i}},}
where $\sigma_i,s_i$ are the $z$ components of the spin-1/2 and spin-1 operators and they take the values
$\sigma_i=\pm 1$ and $s_i=0,\pm 1$, respectively.  The positive valued $J_{AA},J_{AB}=J_{BA}, J_{BB}$  are the ferromagnetic
exchange interactions between the nearest neighbor spins, $D$ is
the crystal field (single ion anisotropy) at a site $i$, $H$ is the
external longitudinal magnetic field. Here $\xi_i$ and $\delta_i$ are the site occupation numbers which can take values $0,1$ and they related to each other by the expression $\xi_i+\delta_i=1$ (i.e. lattice has no vacancy). $\xi_i=1-\delta_i=1$ means that the site i has A type atom and $\xi_i=1-\delta_i=0$ means that the site i has B type atom.  The first summation in Eq.
\re{denk1} is over the nearest-neighbor pairs of spins and the other
summations are over all the lattice sites.

In order to construct EFT equations of the system, let us write Hamiltonian of the system related to the site $0$ by using local fields as

\eq{denk2}{
\mathcal{H}_0= \mathcal{H}_0^{A}+\mathcal{H}_0^{B},} 
where the local fields $ \mathcal{H}_0^{A} , \mathcal{H}_0^{B} $ acts on site $0$ if the site has A/B type atoms, respectively. These local fields represents the spin-spin interactions of the site $0$ ($E_0^{A},E_0^{B}$) as well as interactions with external fields and given as
\eq{denk3}{
\mathcal{H}_0^{A}=-\xi_0\sigma_0\left[\summ{j=1}{z}{\paran{\xi_jJ_{AA} \sigma_j + \delta_j J_{AB} s_j }}+H\right]=-\xi_0\sigma_0\left[E_0^{A}+H\right],} 

\eq{denk4}{
\mathcal{H}_0^{B}=-\delta_0 s_0\left[\summ{\delta=1}{z}{\paran{\xi_jJ_{BA} \sigma_j + \delta_j J_{BB} s_j }}+H\right]-\delta_0\paran{s_0}^2 D =
-\delta_0 s_0\left[E_0^{B}+H\right]-\delta_0\paran{s_0}^2 D.}

The equations related to the order parameters of the system  can be calculated via exact identities
\cite{ref44} which are given by
$$
m_A=\frac{\sanddr{\xi_0\sigma_0}}{\sandd{\xi_0}_r}=\frac{1}{\sandd{\xi_0}_r}\sanddr{\frac{Tr_0\xi_0 \sigma_0 \exp{\paran{-\beta
\mathcal{H}_0^{A}}}}{Tr_0\exp{\paran{-\beta \mathcal{H}_0^{A}}}}},
$$

\eq{denk5}{m_B=\frac{\sanddr{\delta_0s_0}}{\sandd{\delta_0}_r}=\frac{1}{\sandd{\delta_0}_r}\sanddr{\frac{Tr_0 \delta_0 s_0 \exp{\paran{-\beta
\mathcal{H}_0^{B}}}}{Tr_0\exp{\paran{-\beta \mathcal{H}_0^{B}}}}},}

$$
q_B=\frac{\sanddr{\delta_0s_0^2}}{\sandd{\delta_0}_r}=\frac{1}{\sandd{\delta_0}_r}\sanddr{\frac{Tr_0 \delta_0 s_0^2
\exp{\paran{-\beta \mathcal{H}_0^{B}}}}{Tr_0\exp{\paran{-\beta
\mathcal{H}_0^{B}}}}},
$$
 where $Tr_0$ is the partial trace over the site
$0$, $\beta=1/\paran{k_B T}$, $k_B$ is Boltzmann constant
and $T$ is the temperature.  In Eq. \re{denk5},  the
inner brackets stand for the thermal average, while the outer ones
(with subscript $r$) stands for the random configurational average,
which is necessary for including  the effect of the random distribution of the type of A/B atoms on lattice.

By writing Eqs. \re{denk3} and \re{denk4} in Eq. \re{denk5} and performing partial trace operations we can obtain order parameters as

\eq{denk6}{
m_A=\sanddr{f\paran{ E_0^{A}}}, \quad m_B=\sanddr{g\paran{ E_0^{B}}}, \quad q_B=\sanddr{h\paran{ E_0^{B}}},
}
where the functions

$$
f(x)=\tanh{\left[\beta\paran{x+H}\right]}
$$

\eq{denk7}{
g(x)=\frac{2\sinh{\left[\beta\paran{x+H}\right]}}{2\cosh{\left[\beta\paran{x+H}\right]}+\exp{\paran{-\beta
D}}}
}
$$
h(x)=\frac{2\cosh{\left[\beta\paran{x+H}\right]}}{2\cosh{\left[\beta\paran{x+H}\right]}+\exp{\paran{-\beta
D}}}.
$$

We note that, while obtaining Eq. \re{denk6} the following identity has been used
\eq{denk8}{
e^{\xi x}=\xi e^{x}+1-\xi,
} where $x$ is any real number and $\xi=0,1$.  

In a standard differential operator technique \cite{ref45}, Eq. \re{denk6} can be written as

\eq{denk9}{
m_A=\sanddr{e^{E_0^{A}\nabla}}f(x)|_{x=0} \quad m_B=\sanddr{e^{E_0^{B}\nabla}}g(x)|_{x=0}, \quad q_B=\sanddr{e^{E_0^{B}\nabla}}h(x)|_{x=0},
}
where $\nabla$ is the differential operator with respect to  $x$ and  the
effect of the differential operator on an arbitrary function is
given by
\eq{denk10}{\exp{\paran{a\nabla}}F\paran{x}=F\paran{x+a}} with any constant  $a$.

Eq. \re{denk9} can be evaluated by using Van der Waerden identities for spin-1/2 and spin-1 systems \cite{ref46}.  The form of these identities combined with Eq. \re{denk8} can be given by 

\eq{denk11}{
\begin{array}{lcl}
e^{a\xi_j \sigma_j}&=&\xi_j \cosh{\paran{a}}+\xi_j \sigma_j \sinh{\paran{a}}+1-\xi_j,\\
e^{a \delta_j s_j}&=&1 +\delta_j  s_j
\sinh{\paran{a}}+\delta_j \paran{s_j}^2\paran{\cosh{\paran{a}}-1} .
\end{array}
}
Evaluation of Eq. \re{denk9}, using Eqs. \re{denk10} and \re{denk11} creates
multi-spin correlations. Handling of these terms within the decoupling approximation
\cite{ref47} is typical, 
\eq{denk12}{ \sanddr{\ldots \sigma_i \ldots s_j
\ldots \paran{s_k}^2\ldots } =\ldots
\sanddr{\sigma_i}\ldots \sanddr{s_j}\ldots \sanddr{\paran{s_k}^2}\ldots.} If this decoupling approximation procedure is applied to
the expanded forms of the  Eq. \re{denk9}, the resulting equations can be given as

$$
m_A=\summ{p=0}{z}{}\summ{r=0}{z-p}{}\summ{s=0}{p}{}\summ{t=0}{s}{}
C_{prst}f\paran{\paran{z-p-2r} J_{AA}+\paran{s-2t}  J_{AB}},
$$
\eq{denk13}{
m_B=\summ{p=0}{z}{}\summ{r=0}{z-p}{}\summ{s=0}{p}{}\summ{t=0}{s}{}
C_{prst}g\paran{\paran{z-p-2r} J_{BA}+\paran{s-2t}  J_{BB}},}
$$
q_B=\summ{p=0}{z}{}\summ{r=0}{z-p}{}\summ{s=0}{p}{}\summ{t=0}{s}{}
C_{prst}h\paran{\paran{z-p-2r}  J_{BA}+\paran{s-2t}  J_{BB}},
$$
where the coefficients are defined by

\eq{denk14}{
C_{prst}=\komb{z}{p}\komb{z-p}{r}\komb{p}{s}\komb{s}{t}c^{z-p}\paran{1-c}^pm_+^{z-p-r}m_-^{r}\paran{1-q_B}^{p-s}q_+^{s-t}q_-^{t}.
} In Eq. \re{denk14} the terms are given by
\eq{denk15}{
m_+=\frac{1+m_A}{2}, \quad m_-=\frac{1-m_A}{2}, \quad q_+=\frac{q_B+m_B}{2}, \quad q_-=\frac{q_B-m_B}{2}
}
and 
\eq{denk16}{
c=\sandd{\xi_i}_r
} is the concentration of the type A (spin-1/2) atoms.

By solving the system of nonlinear equations given by  Eq.
\re{denk13}  with the coefficients given in \re{denk14}, we
get EFT results for $m_A,m_B,q_B$ and the magnetization of the system is obtained via
\eq{denk17}{
m=cm_A+\paran{1-c}m_B.
}

Linearization of Eq. \re{denk13} in $m_A$ and $m_B$ will give the linear equation system whose solution for the temperature gives the second order critical temperature of the system.




\section{Results and Discussion}\label{results}

We will use scaled (dimensionless) quantities throughout the study
as
\eq{denk18}{ r_{AB}=\frac{J_{AB}}{J_{AA}} =\frac{J_{BA}}{J_{AA}}, r_{BB}=\frac{J_{BB}}{J_{AA}}, d=\frac{D}{J_{AA}},t=\frac{k_BT}{J_{AA}},h=\frac{H}{J_{AA}}. }
The hysteresis loops can be obtained for a given set of Hamiltonian parameters by calculating $m$ according to the procedure given
above, and by  sweeping the longitudinal magnetic field from $-h_0$
to $h_0$ and reverse direction (i.e. $h_0\rightarrow - h_0$). We
study on honeycomb lattice (i.e. $z=3$) within this work. Since we want to mainly focus on the effect of the concentration on the hysteresis behaviors, we set the values of $r_{AA}=r_{AB}=r_{BA}=r_{AB}=1$.

\begin{figure}[h]
\epsfig{file=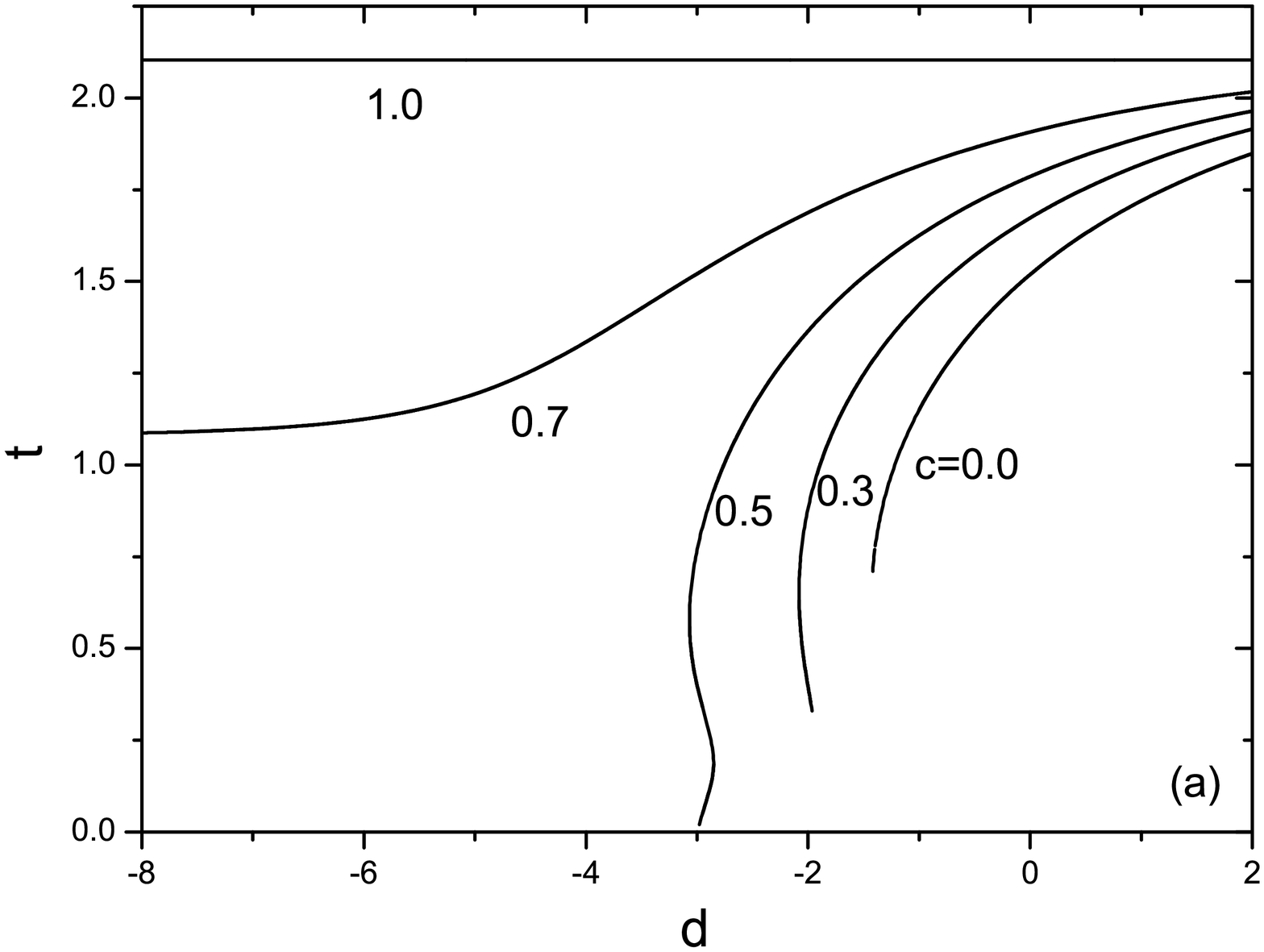, width=8cm}
\epsfig{file=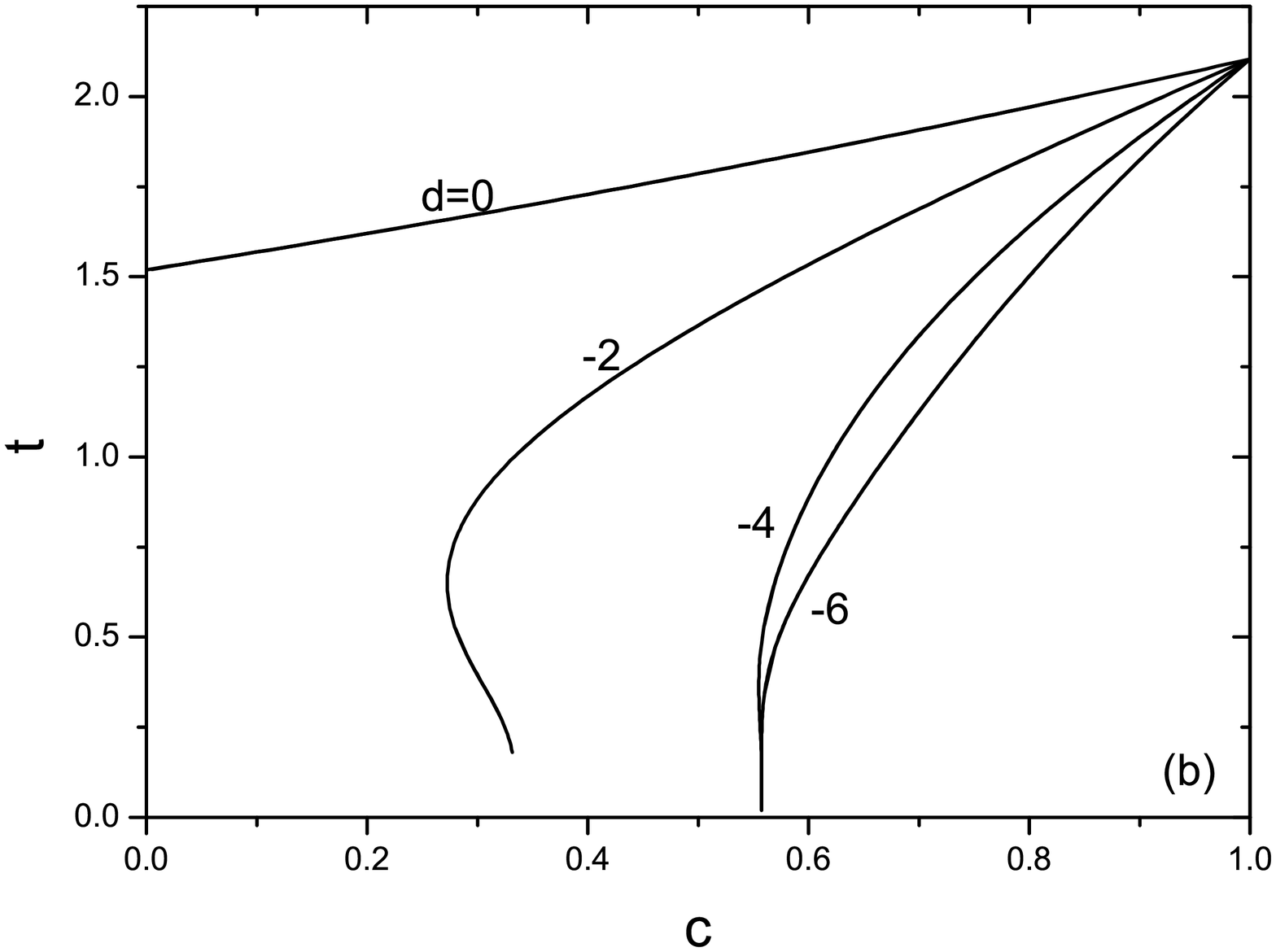, width=8cm}
\caption{Variation of critical temperature of the system (a) with crystal field for concentration values $c=0.0,0.3,0.5,0.7,1.0$ (b) with concentration  for crystal field values of $d=0.0,-2.0,-4.0,-6.0$}\label{sek1}
\end{figure}

The phase diagrams of the system have already been obtained in several works such as within the framework of MFA \cite{ref9,ref12} and  EFT \cite{ref18,ref19,ref22,ref24}. The evolution of the phase diagrams with the concentration in $(t,d)$ plane can be seen in Fig. \ref{sek1} (a). When the value of  $c$ rises, the phase diagrams evolve between the two limiting cases such that $c=0$ (spin-1 Ising model) and  $c=1$ (spin-$1/2$ Ising model). It is well known fact that, $c=0$ curve has first order transition part and tricritical point, which is just the junction of second order and first order parts of the phase diagram.  When $c$ rises, first order portion of the diagram  disappears (compare the curves for $c=0$ and $c=0.3$ in Fig. \ref{sek1} (a)). After then, the phase diagrams evolve to a parallel line to the $d$ axis which is the phase diagram for $c=1$, i.e. spin-$1/2$ Ising model. This value is $t_c=2.104$ which is the critical temperature of the spin-$1/2$ Ising model for the honeycomb lattice within the formulation of EFT. 	 
In Fig. \ref{sek1} (b), the phase diagrams in $(t,c)$ plane for selected values of $d$ can be seen. Note that, the value of $t_c$ for $d=0$ curve at $c=0$ is the critical temperature of the spin-1 Ising model (with zero crystal field), the value at the other side  ($c=1$) is the critical temperature of spin- $1/2$ model. When $d$ takes negative values, disordered phase appears for ground state with lower values of $c$. Note here that, when the value of crystal field lowers, the concentration value which is the border between the ordered and disordered phase at ground state are the same. This value is $c=0.557$.


\begin{figure}[h]
\epsfig{file=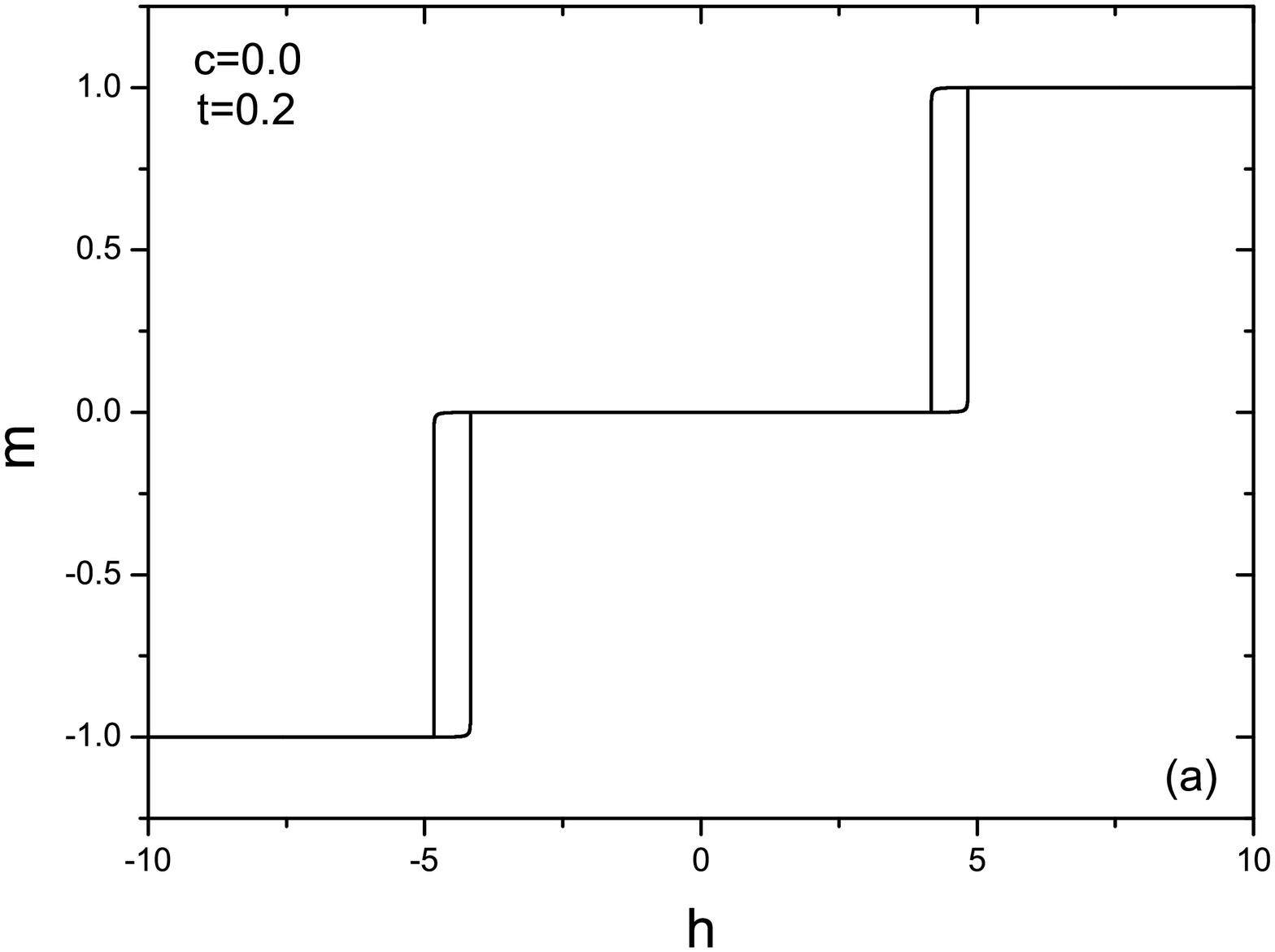, width=7cm}
\epsfig{file=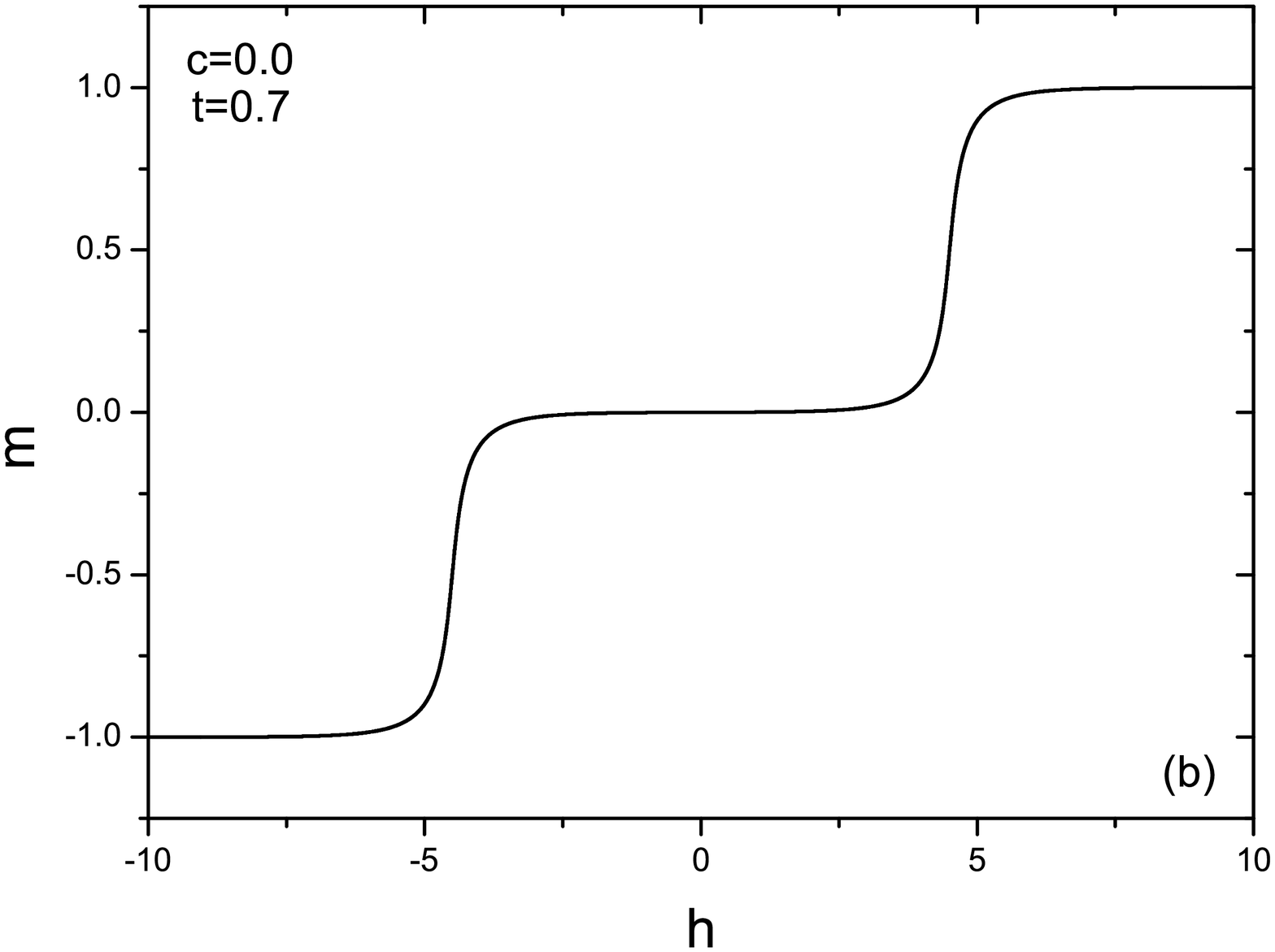, width=7cm}

\epsfig{file=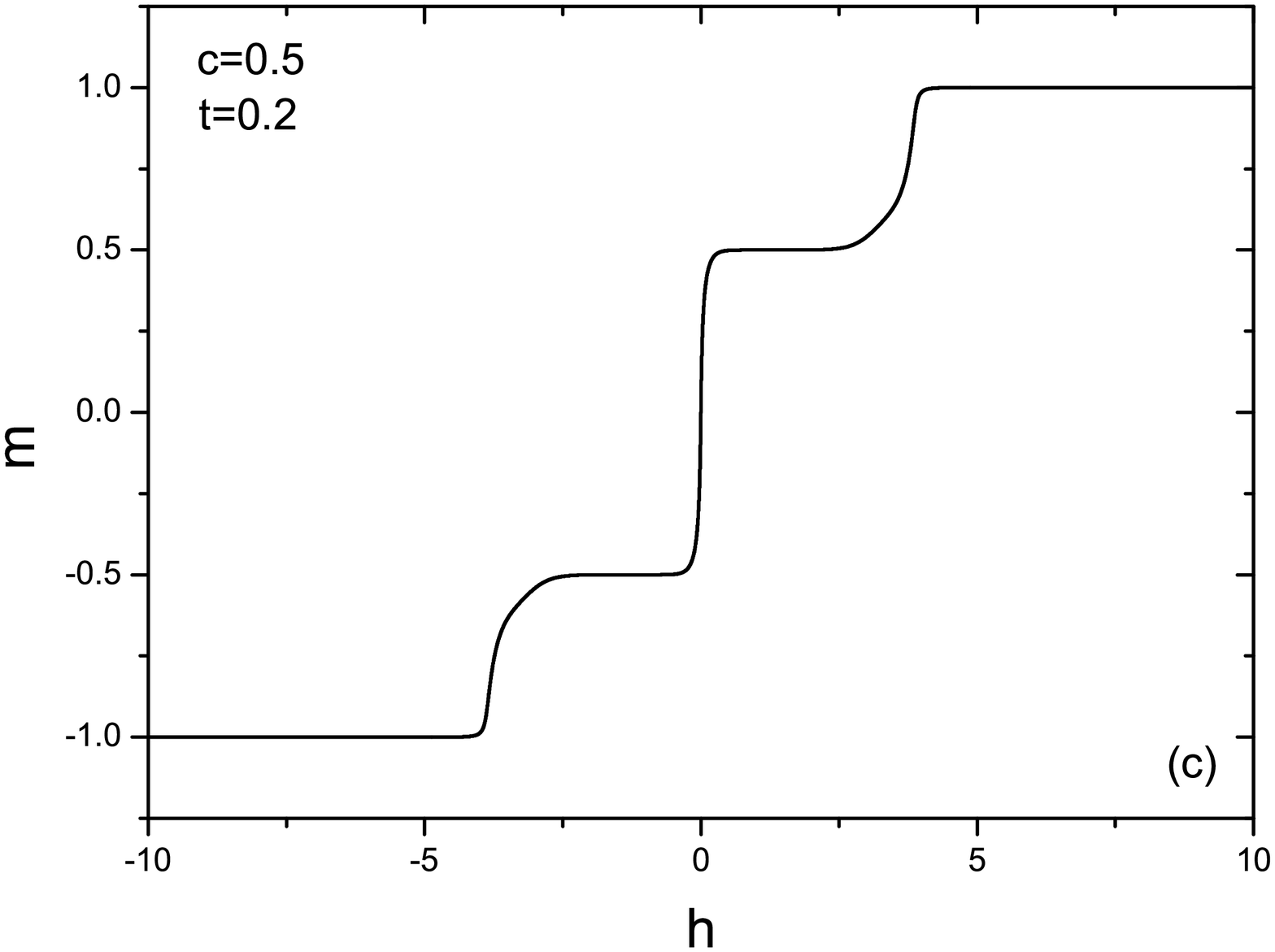, width=7cm}
\epsfig{file=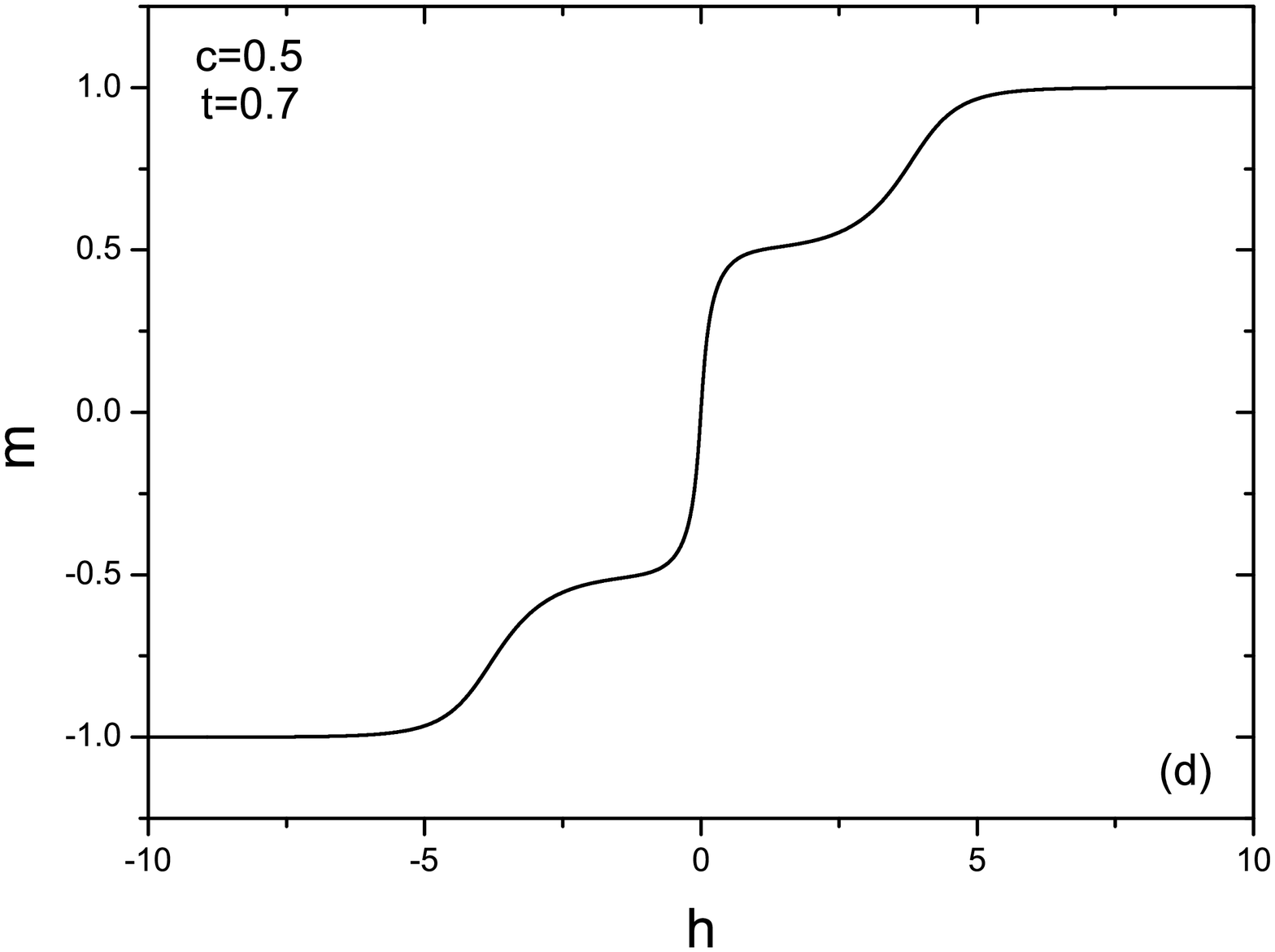, width=7cm}

\epsfig{file=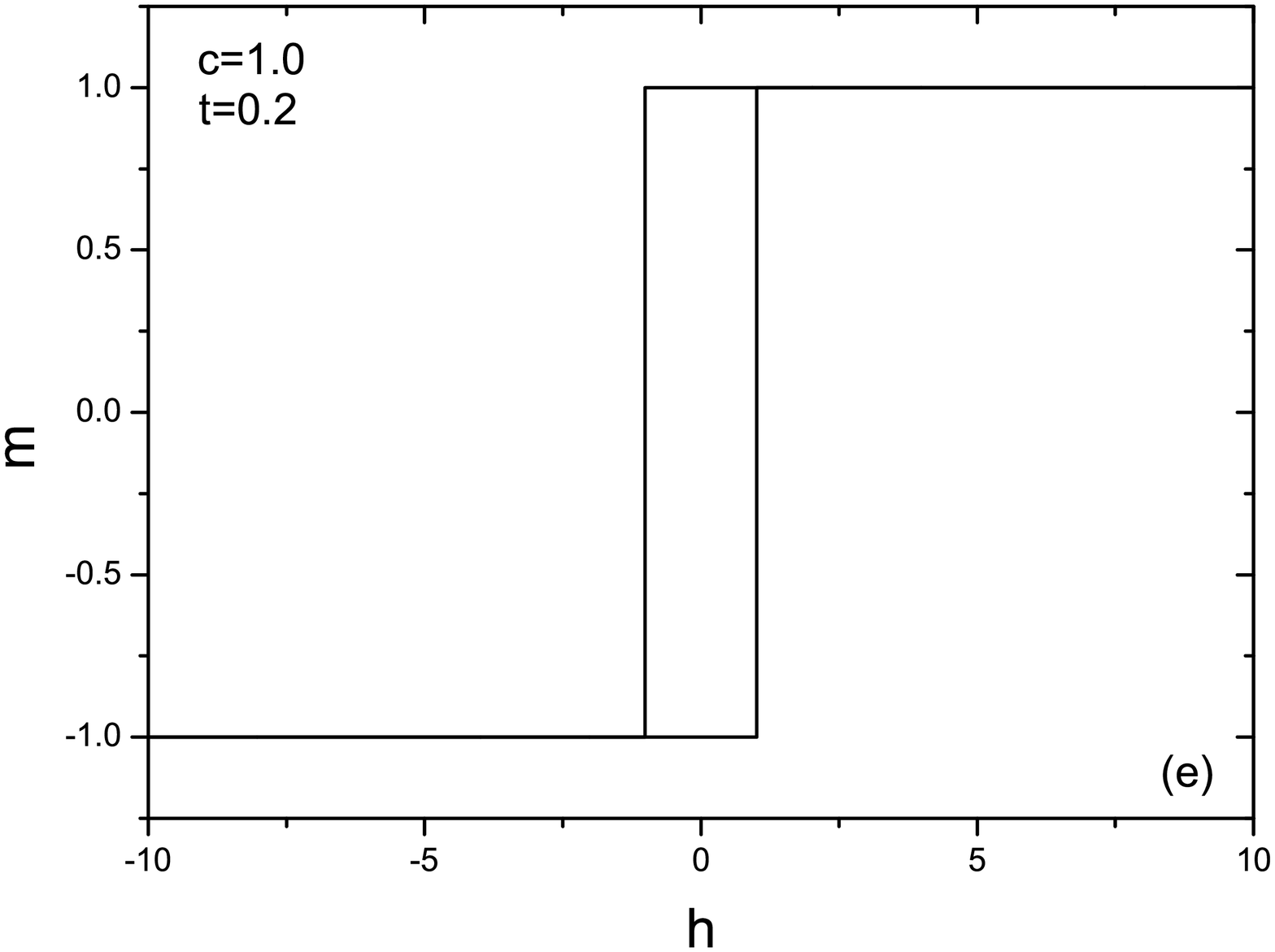, width=7cm}
\epsfig{file=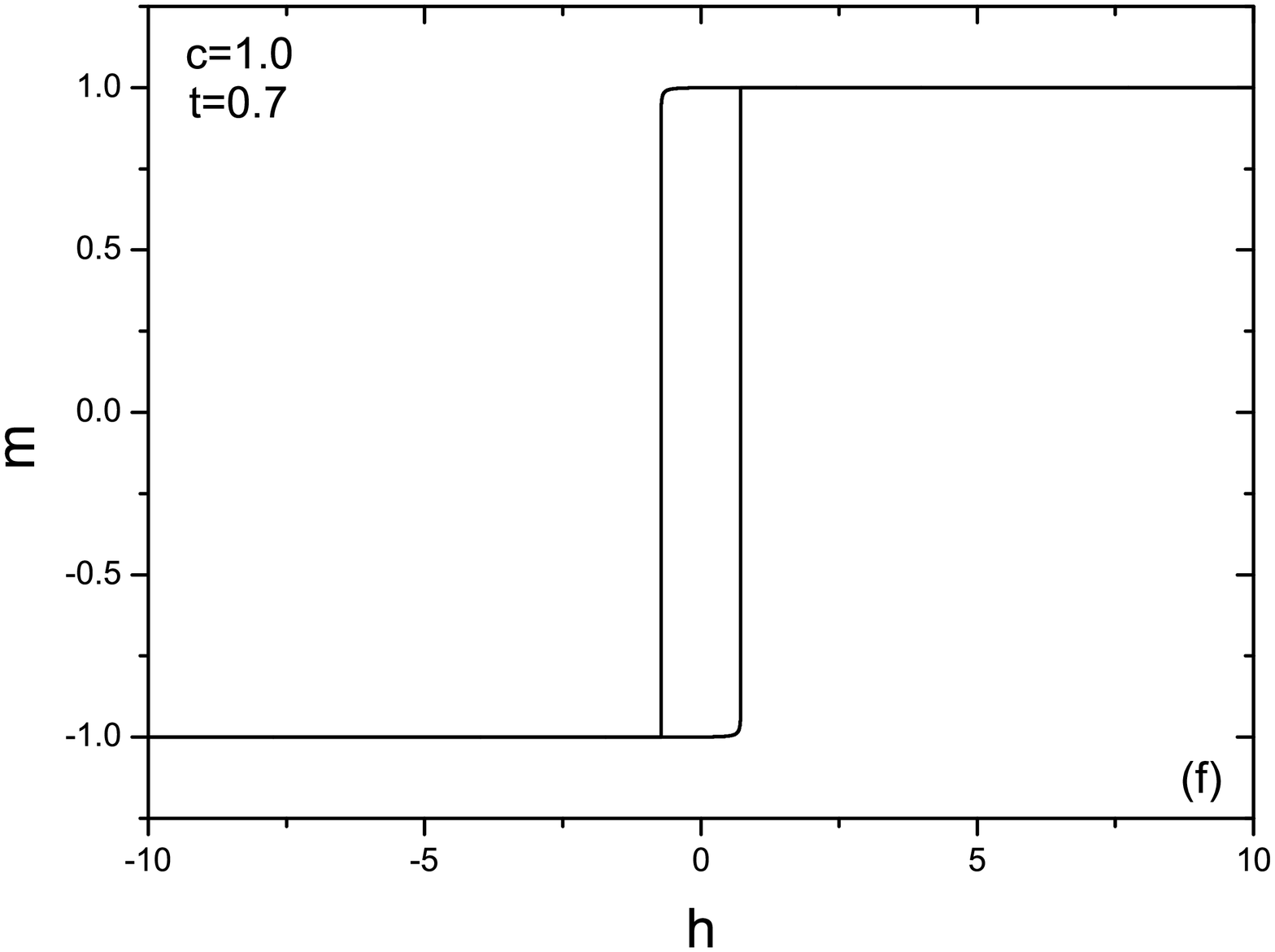, width=7cm}

\caption{Selected hysteresis loops for the spin-$1/2$ spin-1 binary Ising model on honeycomb lattice for the parameter values of $(c,t)$ as (a)
$(0.0,0.2)$, (b) $(0.0,0.7)$, (c) $(0.5,0.2)$, (d)
$(0.5,0.7)$ (e) $(1.0,0.2)$, (f) $(1.0,0.7)$. The value of the crystal field is chosen as $d=-6.0$. }\label{sek2}
\end{figure}

It has been shown that the spin-1 Blume-Capel model has double hysteresis behavior and when the crystal field is diluted, triple hysteresis behavior may occur in a certain range of dilution parameter \cite{ref42}. In this work, it has been shown that the same behavior is valid for the spin-1/2 spin-1 binary Ising model. Before systematic investigation of these behaviors, let us depict representative hysteresis loops of the system. These loops can be seen in In Fig. \ref{sek2}. The crystal field value has been chosen as $d=-6.0$. We can see from Fig. \ref{sek1} (b) that, for the range of $0.0<c<0.557$ system has disordered phase as a ground state, while the system is in an ordered state for the interval $0.557<c<1.0$.  As seen in Fig. \ref{sek2} system can display double hysteresis (DH) behavior (see Fig. \ref{sek2} (a) ), single hysteresis (SH) behavior (see Fig. \ref{sek2} (e) and(f)) and paramagnetic hysteresis (PH) behaviors (see Fig. \ref{sek2} (b), (c) and (d)) according to the values of the Hamiltonian parameters and the temperature.

Rising temperature causes transition to the  PH behavior (compare Figs. \ref{sek2} (b) with (a)), if the amount of the increment on the temperature is enough. On the other hand, rising concentration induce transition from the DH behavior to the SH behavior, for lower values of the temperature (see Figs. \ref{sek2} (a) and (e)). Remind that, rising $c$ results in transition from the disordered phase to the ordered phase (see Fig. \ref{sek1} (b) the curve related to the $d=-6.0$). The reason for the transition from the DH behavior to the SH behavior is similar to the mechanism explained in Ref. \cite{ref42} for the DH behavior in spin-1 BC model. The large negative values of crystal field means that almost all spins of the system are in state $s_i=0$. Rising magnetic field can induce transition from $s_i=0$ state to the $s_i=1$ state. In a similar manner, rising magnetic field in negative direction can induce $s_i=0\rightarrow s_i=-1$ transition (Note that, large value of crystal field dominates the effect of the exchange interaction). These transitions are history dependent, which means that DH behavior. Indeed these plateou type of behavior of the magnetization is a well known behavior \cite{ref22}. When the concentration rises, some sites start to be filled with spin-$1/2$ atoms. Since $s_i=0$ is not allowed state for these atoms, the double hysteresis behavior is weakened. For the value of $c=0.5$ (Fig. \ref{sek2} (c)) half of the lattice sites are occupied by spin-1 atoms while remaining sites have spin-$1/2$ atoms. The ground state is disordered (see Fig. \ref{sek1} (a) or (b)). This disordered state comes from the occupied $s_i=0$ state by spin-$1$ atoms and occupied $s=\pm 1$ states randomly by spin-$1/2$ atoms. Note that since the distribution of spin-$1/2$ atoms are random, exchange interaction between spin-$1/2$ atoms could not align all of the magnetic moments of these atoms in the same direction. Rising magnetic field can cause transition from zero magnetization to value of $0.5$. This value comes from the parallel aligned magnetic moments of spin-$1/2$ atoms to the magnetic field. Since absolute value of the crystal field is greater than the magnetic field, spin-1 atoms could not abandon the $s_i=0$ states. If magnetic field rises further, spin-1 atoms display transition from the  $s_i=0$ states to the $s_i=1$, which is parallel direction to the magnetic field. Same reasoning holds for the rising magnetic field in negative direction. At last, when the concentration is $c=1$, the system consist of only spin-$1/2$  atoms. $s_i=0$ state is not allowed for the atoms then usual SH behavior appears (see Fig. \ref{sek2} (e)). Rising temperature means rising thermal fluctuations and this causes shrinking behavior (or disappearing behavior, if the temperature is high enough) of the hysteresis loops.

\begin{figure}[h]
\epsfig{file=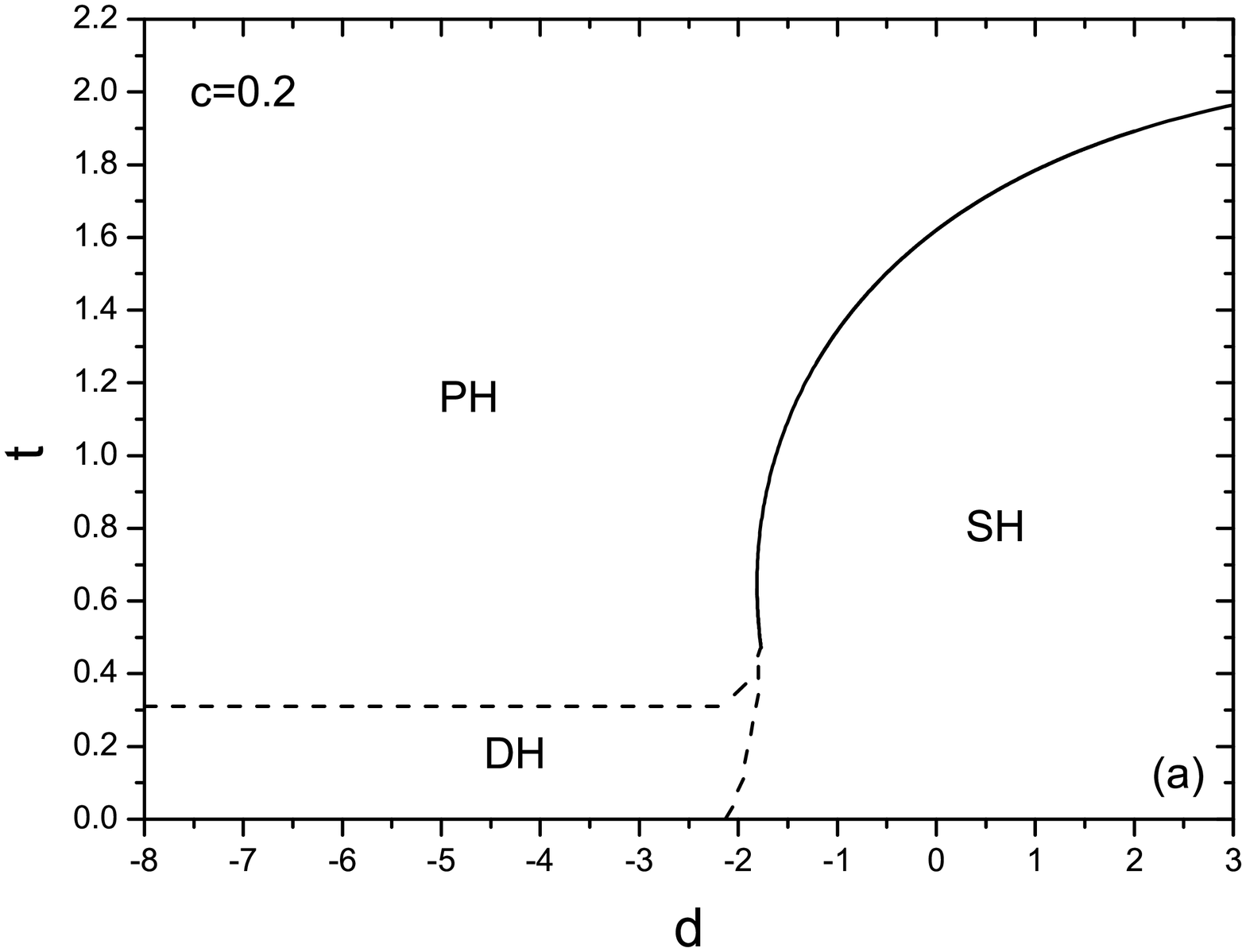, width=7cm}
\epsfig{file=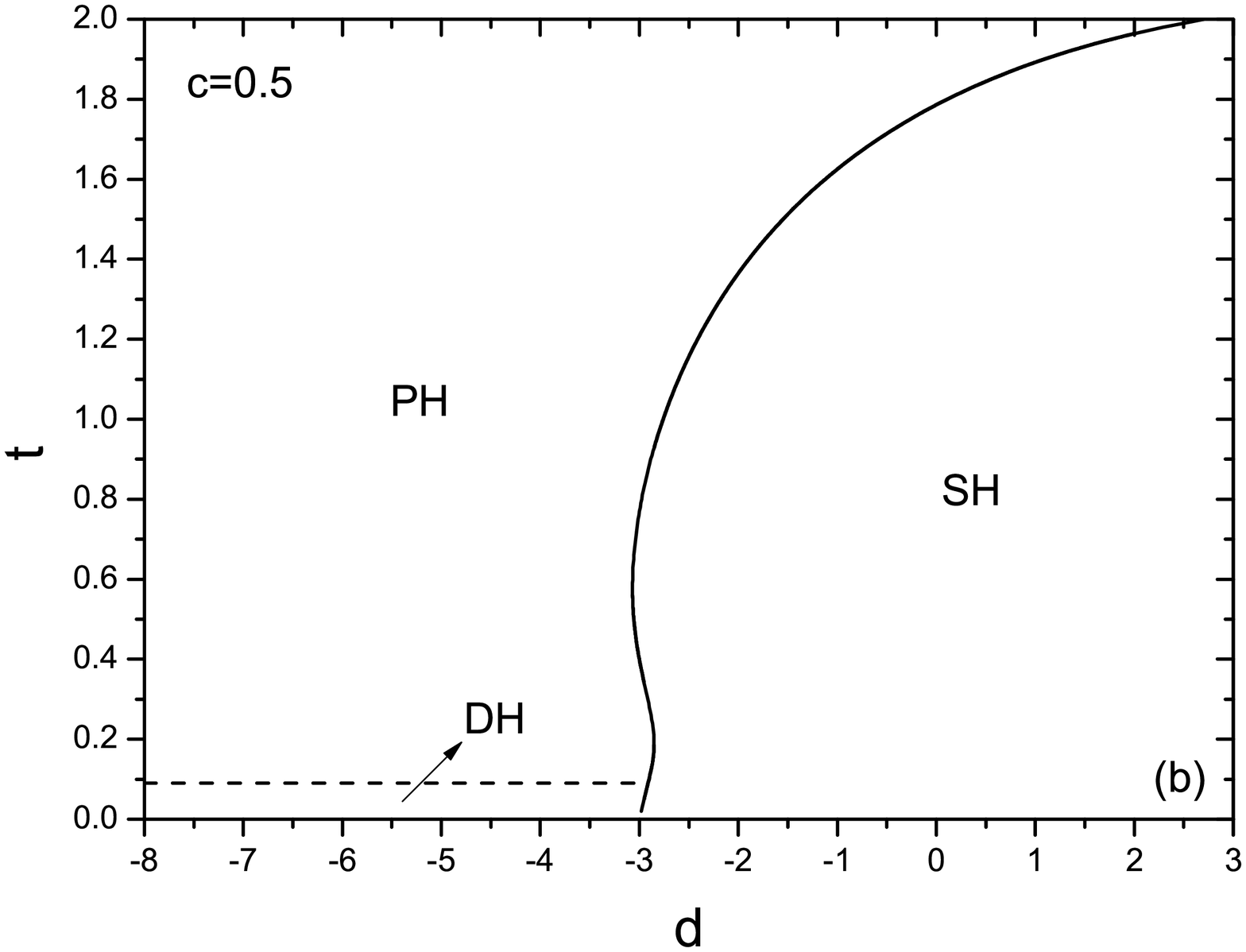, width=7cm}

\caption{Regions that have different hysteresis characteristics in $(t,d)$ plane for selected values of (a) $c=0.2$ and (b) $c=0.5$. The crystal field value is set to $d=-6.0$ which represent  large negative value of the crystal field. Note that, solid lines are also second order phase diagrams of the system. }\label{sek3}
\end{figure}

We can determine the regions that have different hysteresis characteristics on $(t,d)$ plane. In Fig. \ref{sek3} we depict these regions for a large negative value of the crystal field ($d=-6.0$). When the concentration is $c=0.2$ we can see from Fig. \ref{sek3} (a) that, DH region covers the  disordered region of the system, which is restricted by the temperature. After a certain value of the temperature, DH behavior is replaced by PH behavior. It is trivial that, ordered region in $(t,d)$ plane is also covered by SH behavior. Regardless of the value of concentration, the border between the SH and other behaviors is also phase diagram of the system. Border between the DH and PH behavior is almost a straight line, which is parallel to the axis $d$ at a specific value of temperature.  By comparing Fig. \ref{sek3} (a) and (b) we can conclude that, the value of $t$ that is the height of the border between the DH and PH in $(t,d)$ plane regularly decreases with increasing concentration. This border disappears at the concentration value of $c=0.557$ which separates ordered and disordered phases for large negative values of the crystal field at low temperatures (i.e. intersection point of the the curve for $d=-6.0$ with $c$ axis in Fig. \ref{sek2} (b)). 

The border that separates the DH and PH behavior is related to the thermal agitations in the system. It was concluded that, DH behavior comes from the fact that, rising magnetic field can induce the transition from occupied states $s_i=0$ at large negative values of the crystal field. Rising temperature could destroy this occupancy of states $s_i=0$ due to the thermal agitations. Since rising $c$ means that, lowering the percentage of spin-1 atoms, it will be easier to destroy DH behavior when $c$ increases.

Apart from these DH behaviors, the relation between the concentration and the quantities such as hysteresis loop area (HLA), remanent magnetization (RM) and coercive field  (CF) is important. HLA is simply defined as the area covered by hysteresis loop in $(m,h)$ plane and it 
corresponds to the energy loss due to the hysteresis. The RM is residual magnetization  in the system after an external magnetic field is
removed and CF is defined as the intensity of the external magnetic field needed to change the sign of the magnetization. At the crystal field value of $d=0$ the critical temperature of the $c=1$ is higher than  $c=0$. This also means that, the temperature value of which HLA drops to zero of $c=1$ is higher than the $c=0$.Thus it is expected that, the CF value for $c=1$ is larger than $c=0$ since, the system consisting of only spin-$1/2$ atoms ($c=1$) is more resistive to the change of magnetic field than the  system consisting of only spin-1 atoms ($c=0$).

\begin{figure}[h]
\epsfig{file=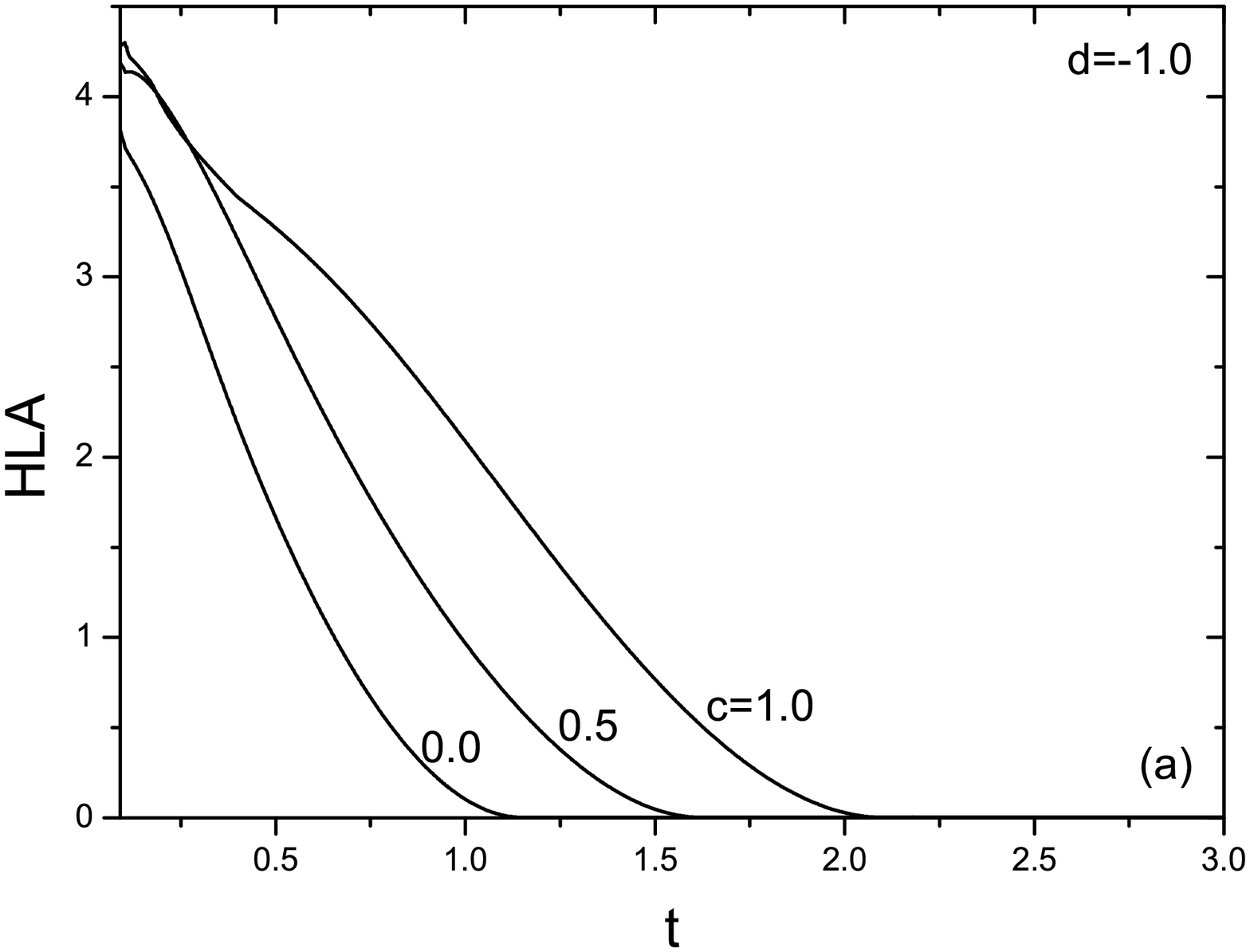, width=5cm}
\epsfig{file=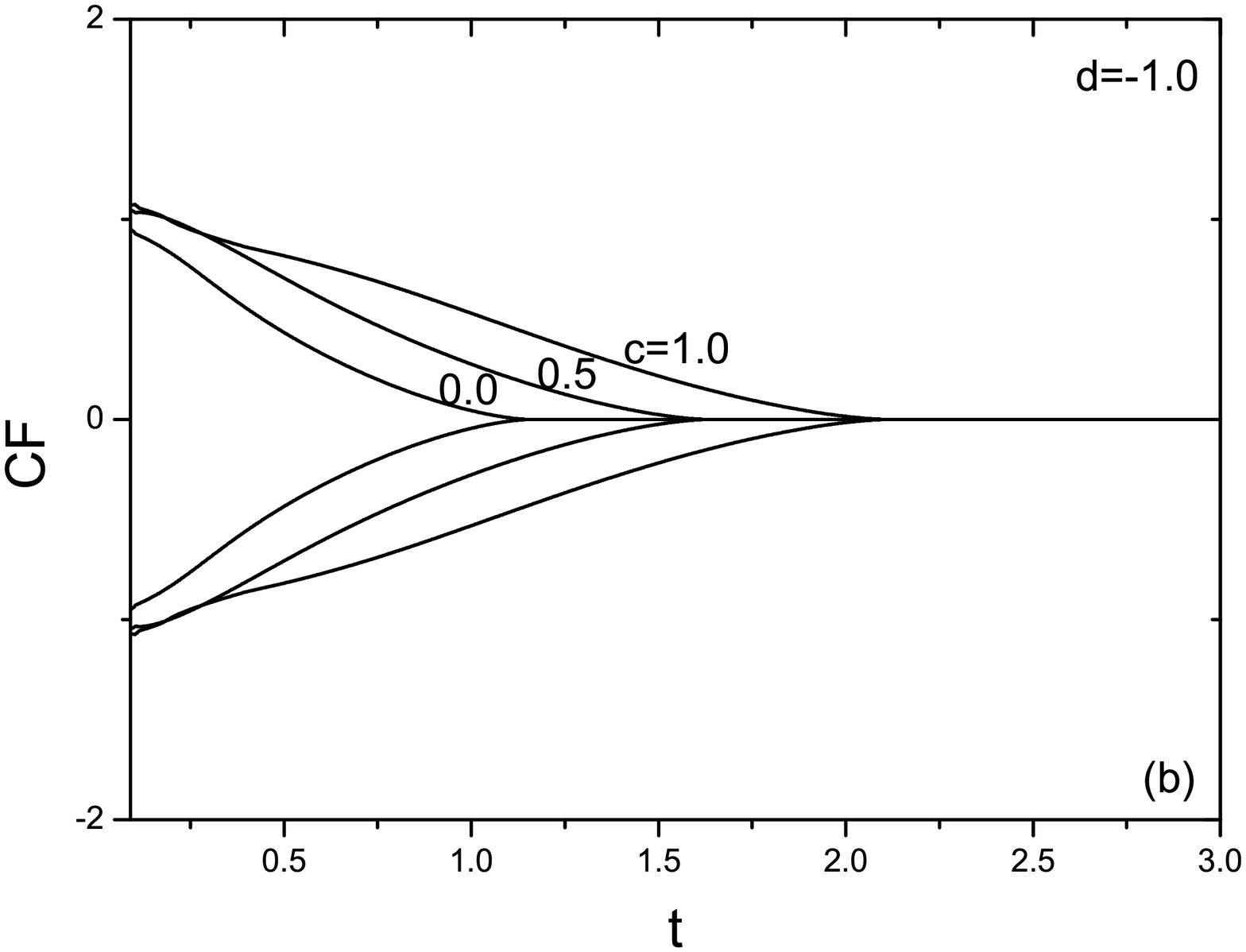, width=5cm}
\epsfig{file=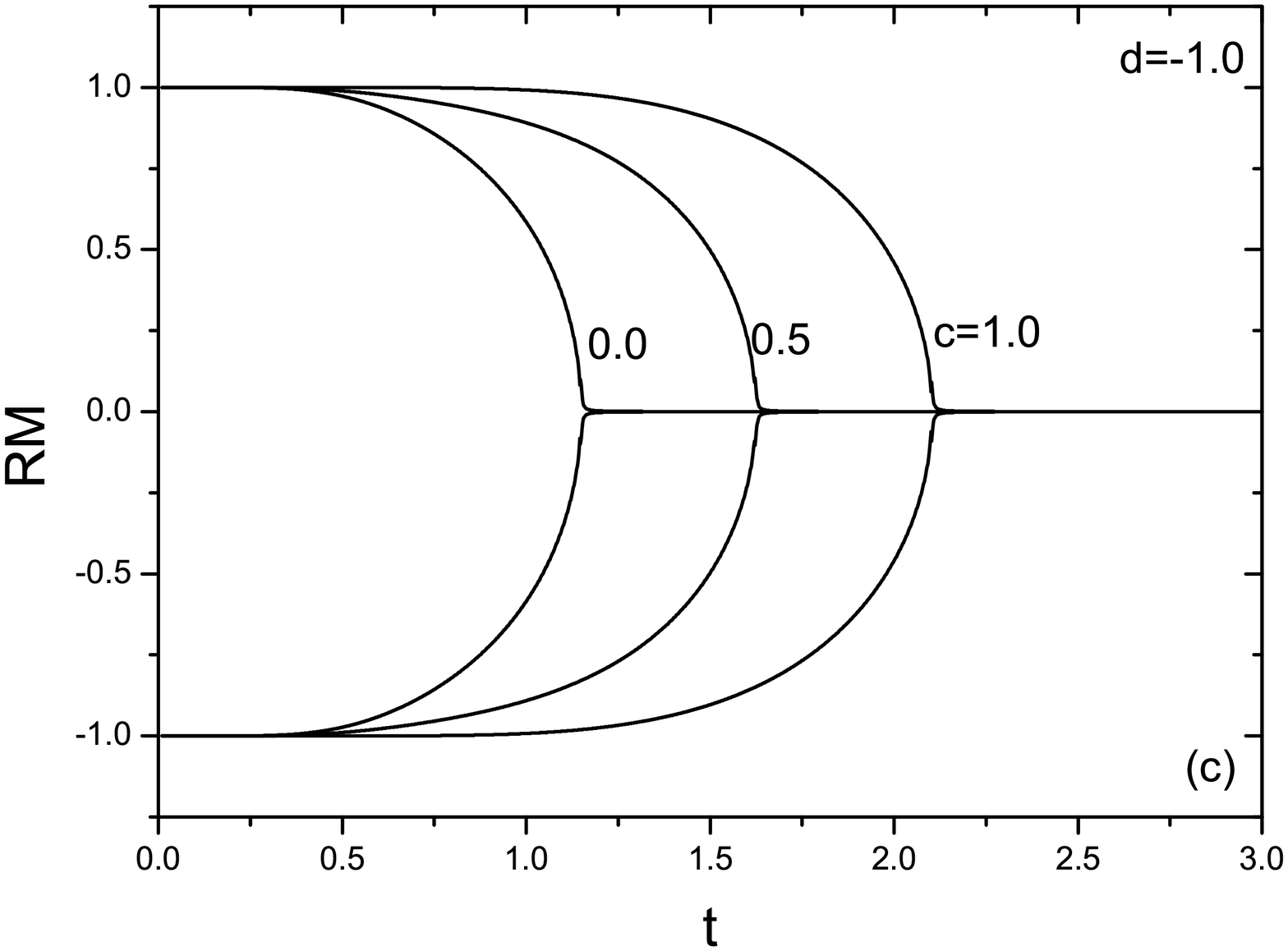, width=5cm}

\epsfig{file=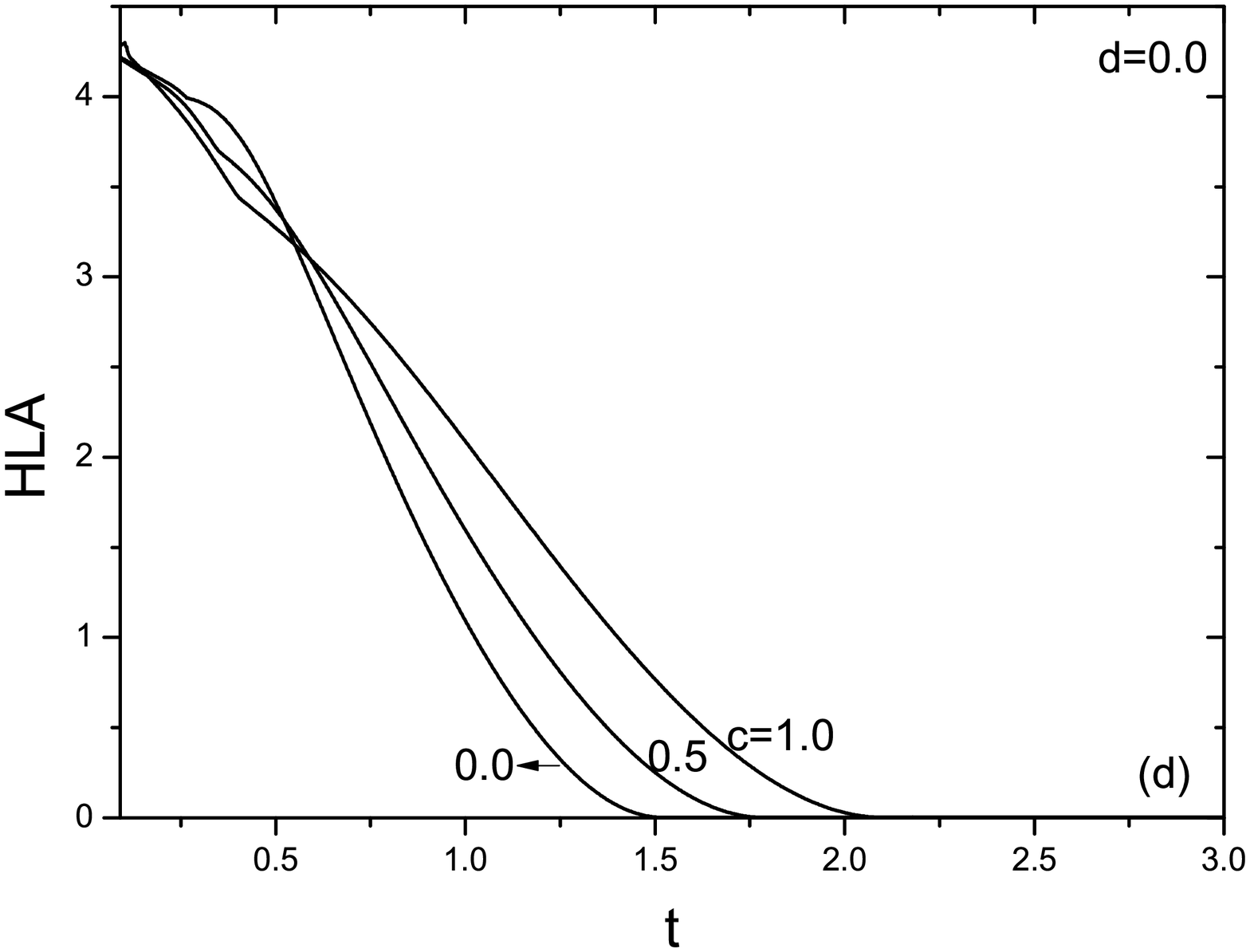, width=5cm}
\epsfig{file=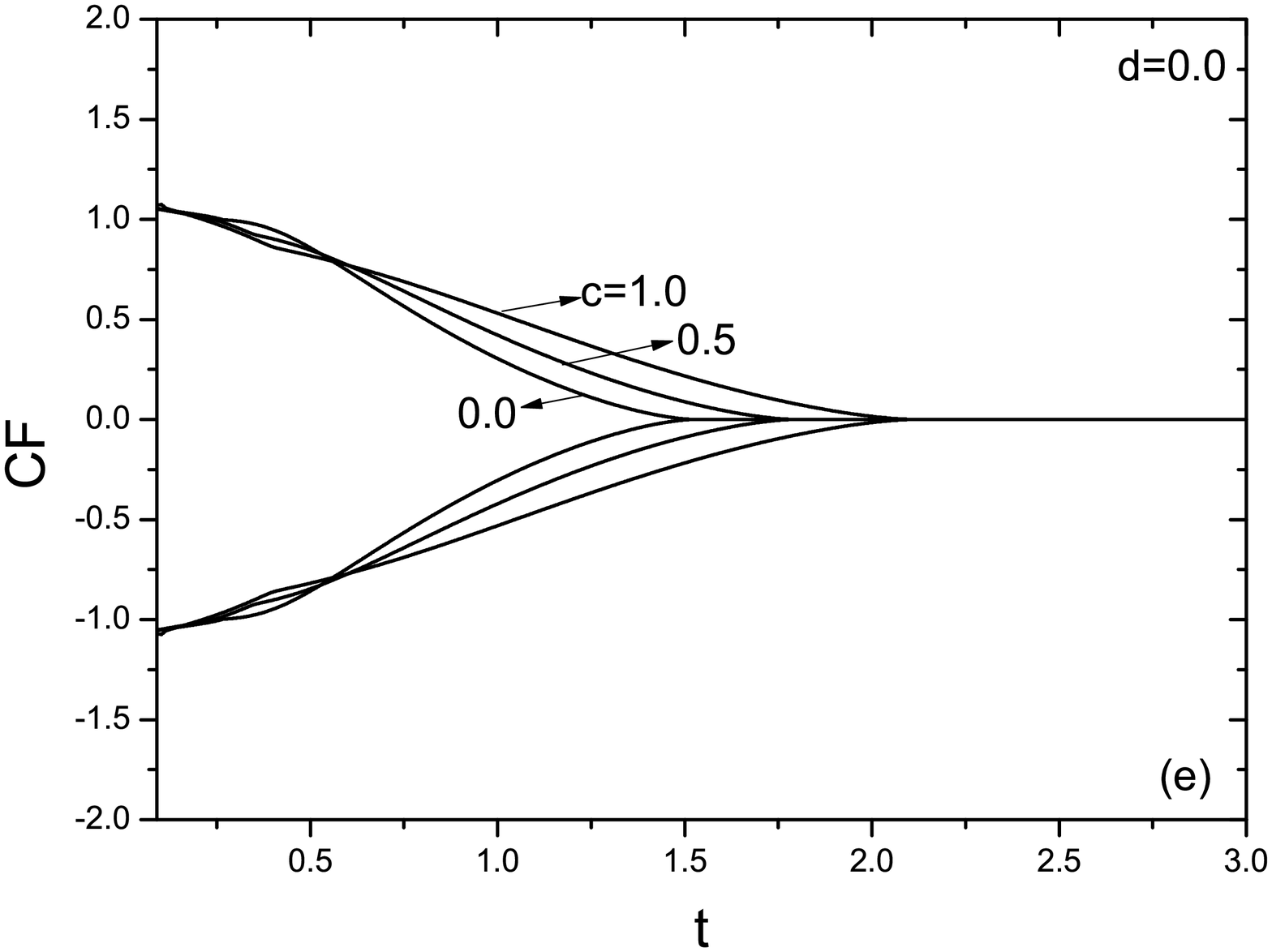, width=5cm}
\epsfig{file=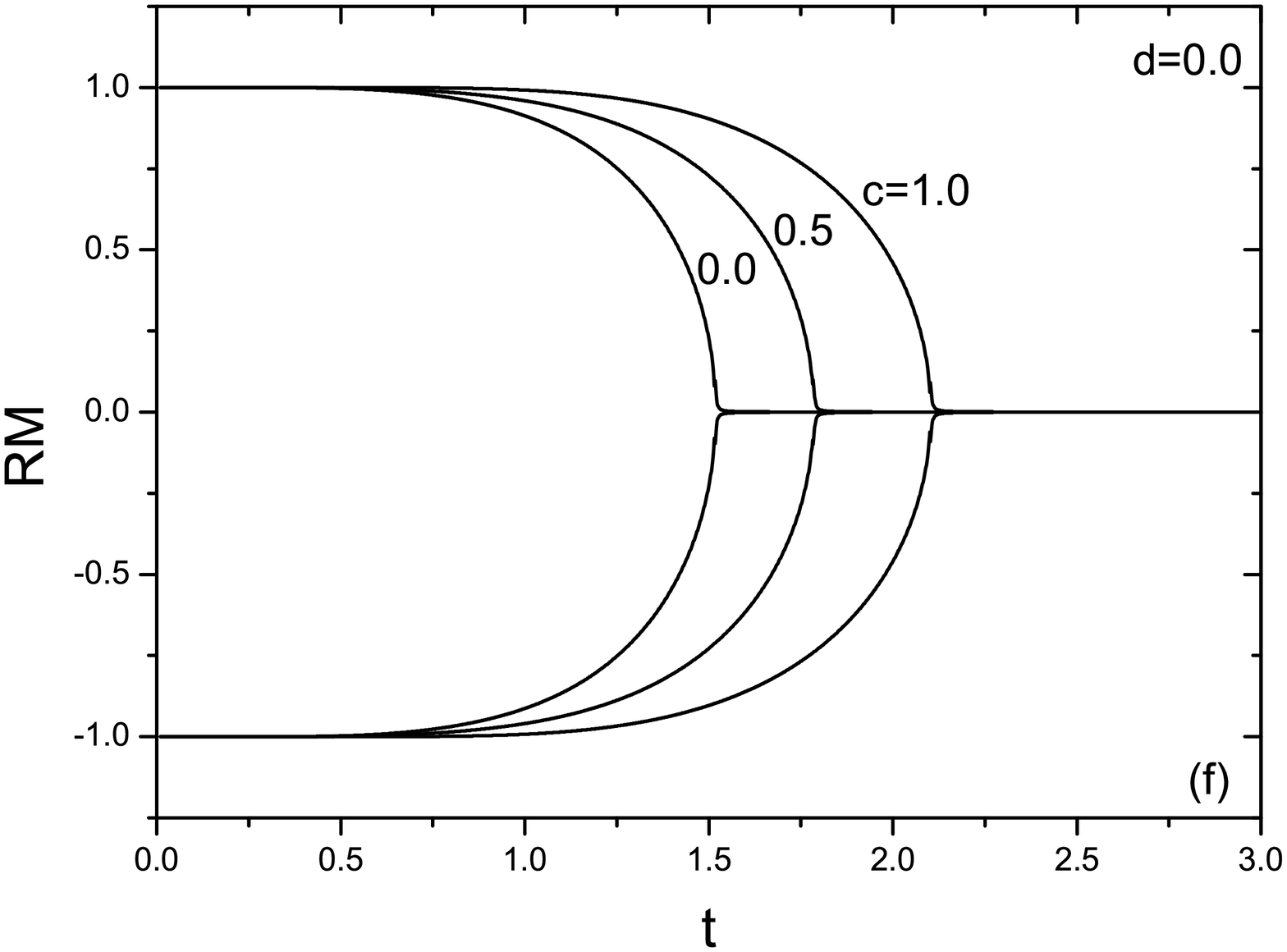, width=5cm}

\epsfig{file=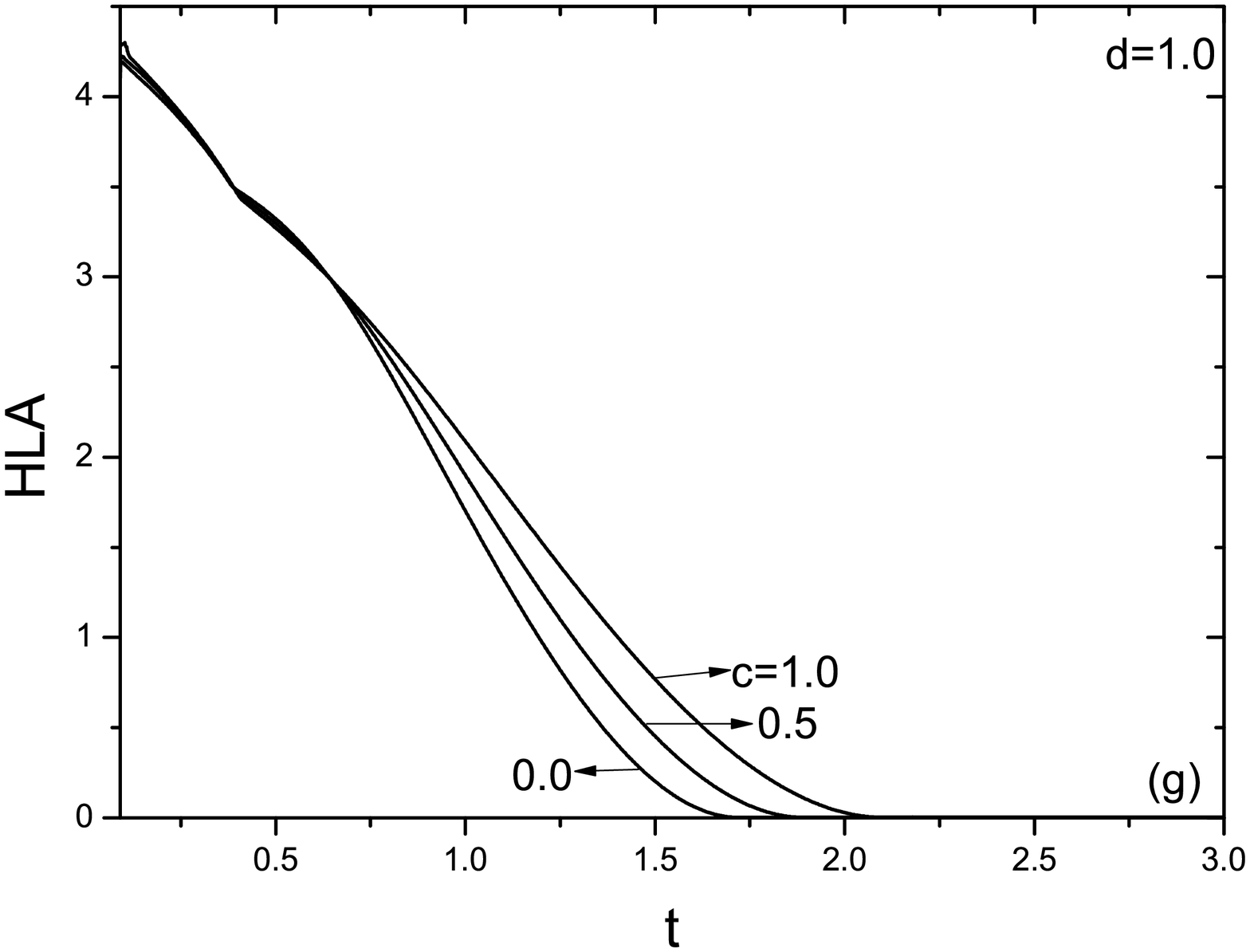, width=5cm}
\epsfig{file=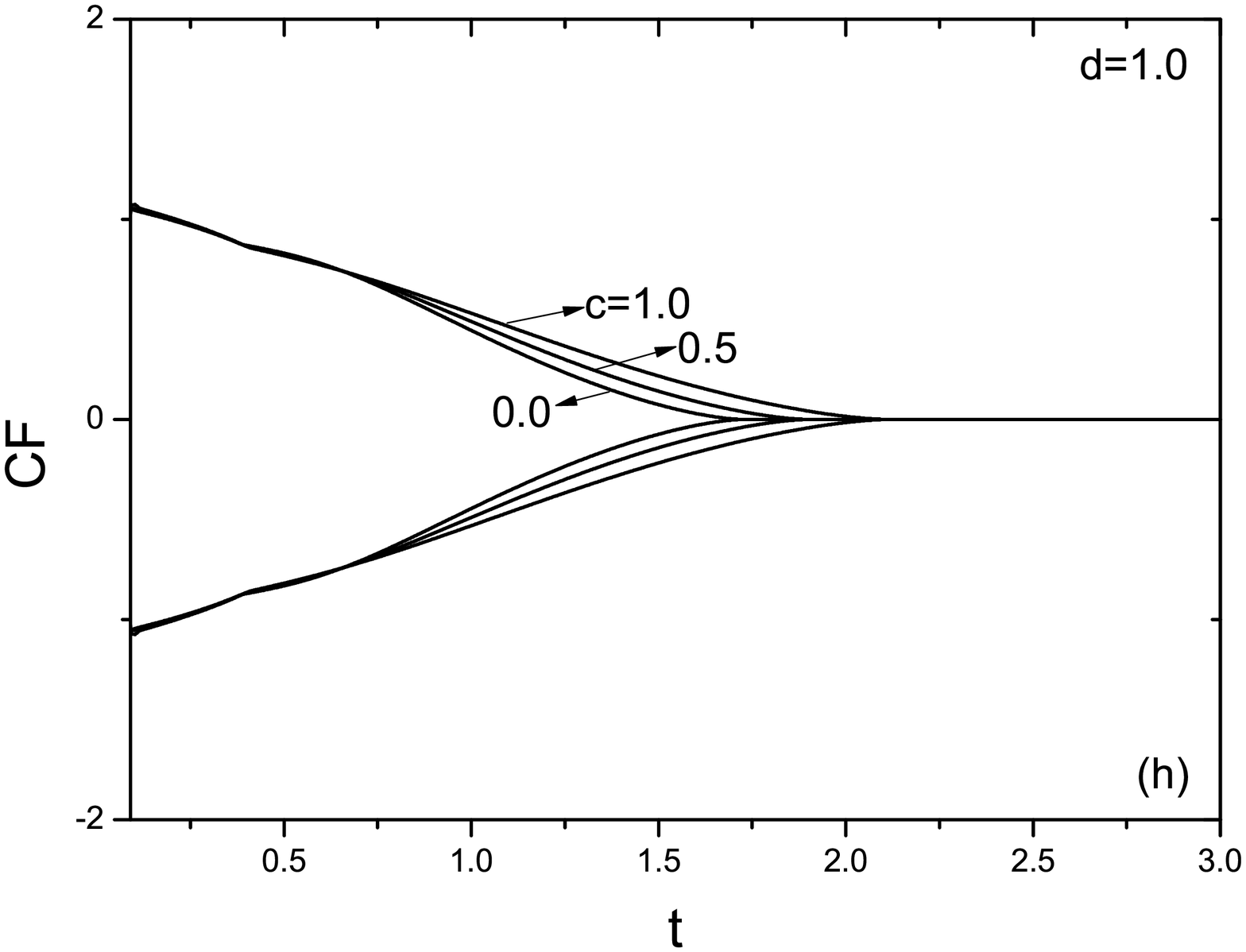, width=5cm}
\epsfig{file=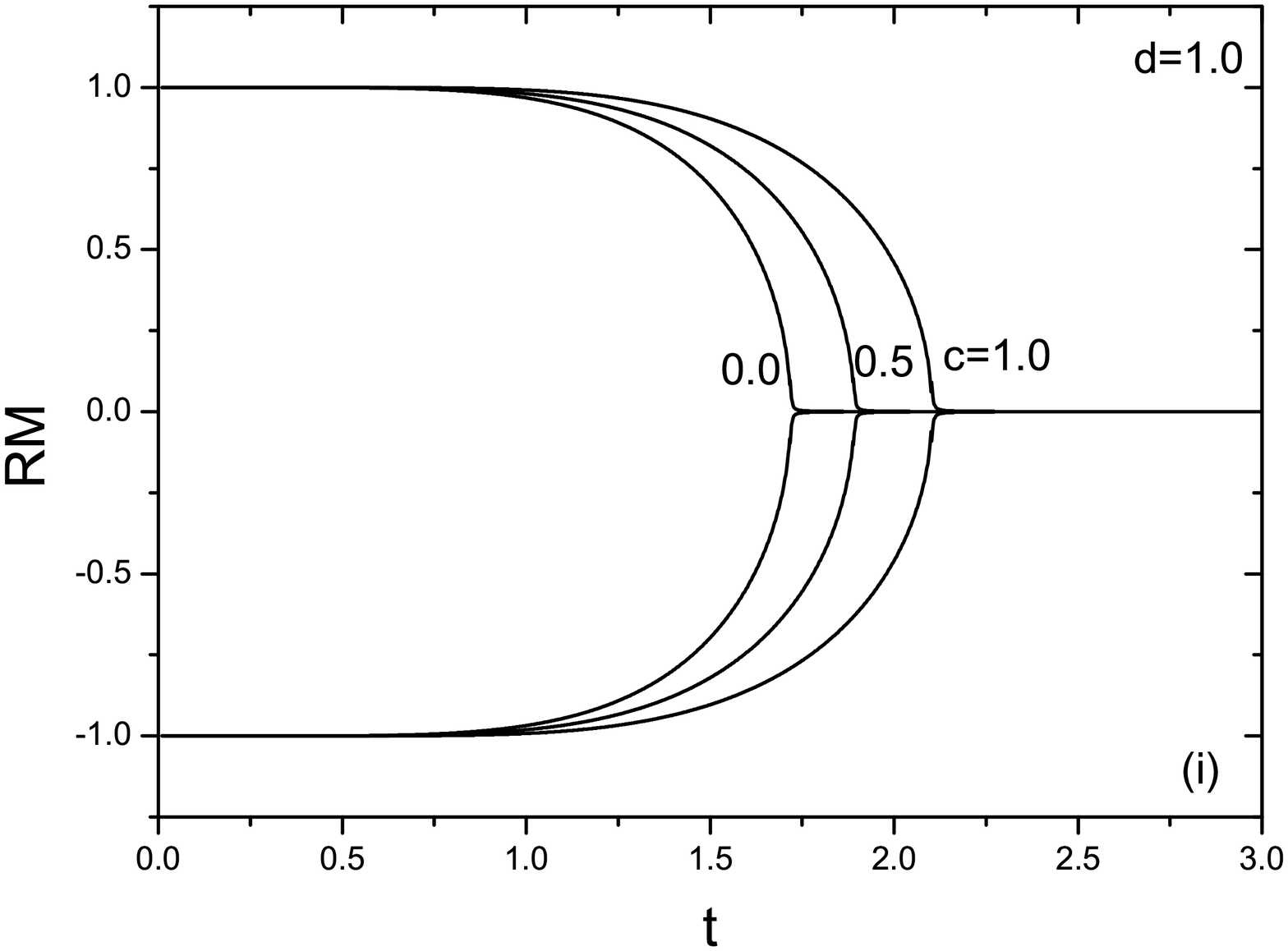, width=5cm}

\caption{Variation of HLA, CF and RM with the temperature for selected values of $d=-1.0,0.0,1.0$ and $c=0.0,0.5,1.0$. }\label{sek4}
\end{figure}

The variation of these quantities with the temperature can be seen in Fig. \ref{sek4} for selected values of $d=-1.0,0.0,1.0$ and $c=0.0,0.5,1.0$. 
As we can see from Figs. \ref{sek4} (a), (d) and (g) that, the value of temperature that HLA drops to zero increases with rising concentration, as expected. Same reasoning holds also for CF (see Figs. \ref{sek4} (b), (e) and (h) ) and RM (see Figs. \ref{sek4} (c), (f) and (i) ).   This type of shrinking behavior of the HLA with rising temperature has been observed in disordered $FeAL$ alloys theoretically with first principle calculations and MC simulation \cite{ref48}.

Again as expected, in most  portions of the graphs, $c=1$ curve lie above of the $c=0$ curve. The difference between these curves decline as $d$ increases (compare Fig. \ref{sek4} (g) and (a)). This is expected since, rising $d$ dictates the magnetic moments of the spin-1 atoms oriented as spin-$1/2$ atoms, i.e. with rising $d$ $s_i=0$ state is not accessible for the spin-1 atoms. In other words, since the limit $d\rightarrow \infty$ in spin-1 model is spin-$1/2$ model, it is expected that, when $d$ rises, the differences between the $c=0$ and $c=1$ in binary alloy system to disappear.   

But as seen in Fig. \ref{sek4} (d), for specific interval of temperature this relation becomes reverse. In other words, HLA (or CF) lowers when $c$ rises in specific interval. If the temperature further increases, relation between the HLA (or CF) and rising concentration gets usual, when $c$ increases, HLA increases. 
This abnormal behavior has to be verified by more accurate methods such as MC simulations.

\section{Conclusion}\label{conclusion}

Hysteresis behaviors of the binary alloy system represented by $A_c B_{1-c}$ is investigated within the framework of EFT. The system consist of type A atoms (spin-$1$) with concentration $c$ and type B atoms (spin-$1/2$) with concentration $1-c$. 

After the phase diagrams given in $(t,d)$ and $(t,c)$ planes, different type of hysteresis behavior examples are given as SH, PH and DH behaviors. There is no new information about the SH and PH behaviors; when the system is in the ferromagnetic region it displays SH behavior, while PH behavior exist in paramagnetic phase. But in some region of the paramagnetic phase, the system could display DH behavior, which is observed for large negative values of the crystal field and low temperatures. Changing of DH behavior to other behaviors with concentration and temperature has been discussed. It has been observed that, binary alloy system could display DH behavior in region $0<c<0.557$. 

Besides, the quantities that  characterize the hysteresis loops has been investigated, namely HLA, CF and RM. Since spin-$1/2$ system is more resistive to the change of magnetic field, the expected result seen in the variations of the HLA, CF and RM with the temperature: $c=1$ curves drop to zero at larger temperatures than the curves related to the  $c=0$. But abnormal behaviors have been detected at lower temperatures, such as when the concentration rises then HLA (and CF) of the system lowers, although $c=1$ curves lie above the other curves at other temperatures. This abnormal behaviors at lower temperatures has to be verified by more sophisticated methods such as MC simulations. We hope that the results
obtained in this work may be beneficial form both theoretical and
experimental points of view.


\end{document}